\documentclass[journal]{IEEEtran}

\usepackage{draftwatermark}\SetWatermarkScale{1.6}

\PassOptionsToPackage{hyphens}{url}\usepackage{hyperref}
\usepackage{amsmath}
\usepackage{graphicx}

\begin{document}

\title{Using Back-Scattering to Enhance Efficiency in Neutron Detectors}

\author{ \IEEEauthorblockN{T.~Kittelmann\IEEEauthorrefmark{1},
    E.~Klinkby\IEEEauthorrefmark{2}\IEEEauthorrefmark{1},
    X.~X.~Cai\IEEEauthorrefmark{2}\IEEEauthorrefmark{1},
    K.~Kanaki\IEEEauthorrefmark{1},
    C.~P.~Cooper-Jensen\IEEEauthorrefmark{1}\IEEEauthorrefmark{3},
    R~Hall-Wilton\IEEEauthorrefmark{1}\IEEEauthorrefmark{4}}
  \\[1ex]\IEEEauthorblockA{\IEEEauthorrefmark{1}European Spallation Source ESS
    AB, Sweden} \\\IEEEauthorblockA{\IEEEauthorrefmark{2}DTU Nutech, Technical
    University of Denmark, Denmark}
  \\\IEEEauthorblockA{\IEEEauthorrefmark{3}Department of Physics and Astronomy,
    Uppsala University, Sweden}
  \\\IEEEauthorblockA{\IEEEauthorrefmark{4}Department of Electronics Design,
    Mid-Sweden University, Sweden}%
  \thanks{Corresponding author is T.\ Kittelmann. Reachable by electronic mail at: \texttt{thomas.kittelmann@esss.se}.}}

\markboth{IEEE Transactions in Nuclear Science,~Vol.~XX, No.~XX, Month~201X}%
{Kittelmann \MakeLowercase{\textit{et al.}}: Using Back-Scattering to Enhance Efficiency in Neutron Detectors}

\maketitle

\begin{abstract}
  The principle of using strongly scattering materials to recover efficiency in
  neutron detectors, via back-scattering of unconverted thermal neutrons, is
  discussed in general. Feasibility of the method is illustrated through
  \texttt{Geant4}-based simulations of a specific setup involving a
  moderator-like material placed behind a single layered boron-10 thin film
  gaseous detector.
\end{abstract}

\begin{IEEEkeywords}
Neutron detectors, Monte Carlo simulations, Geant4, Back-scattering, Boron-10,
Polyethylene
\end{IEEEkeywords}

\IEEEpeerreviewmaketitle

\section{Introduction}

\IEEEPARstart{T}{he} ongoing construction of the European Spallation
Source~\cite{esscdr,esstdr} has initiated significant development of novel
neutronic technologies in the past 5 years.  The performance requirements for
neutron instruments at the European Spallation Source and in particular, the
unprecedented cold and thermal neutron brightness and flux expected from the
source is very challenging for detector technologies currently available.  The
designs for neutron detectors presently in operation at neutron scattering
facilities have seen only incremental improvements over the past decade and are
reaching fundamental performance limits; this has made research into alternative
neutron detectors very topical.

Detection of neutrons with subelectronvolt kinetic energies must necessarily
proceed through destructive nuclear processes in which energetic secondaries are
released and detected themselves. Only a few stable isotopes such as $^3$He,
$^{10}$B, $^6$Li, $^{155,157}$Gd and $^{235}$U have significant cross-sections
for such \emph{conversions}, and detector systems must therefore contain such
materials as well as incorporate capabilities to detect the resulting
secondaries. The dominant detector choice has historically been gaseous $^3$He
detectors, based on the high cross-section process
$\text{n}+^{3\!\!}\text{He}\to^{3\!\!\!\!}\text{H}+p$. However, due to increased
demand and decreased supply, $^3$He will be unavailable in the future for all
but the smallest detectors~\cite{he3crisis1,he3crisis2}. Thus, an extensive
international R\&D programme is currently under
way~\cite{zeitelhack2012,icnd_website} in order to develop efficient and
cost-effective detectors based on other isotopes, and it is expected that
detectors for ESS will be based, as determined by the requirements of a given
instrument, upon technologies ranging from choices such as $^6$Li-containing
scintillator detectors and Gd-based GEM detectors, to gaseous detectors lined
with thin films containing $^{10}$B~\cite{kirstein2014}.

In particular when it comes to instruments where detectors are required to, at a
reasonable cost, cover large areas with detectors offering high rate
capabilities, modest resolutions and modest to high efficiencies, the perhaps
most promising candidate is gaseous detectors surrounded by solid converters in
the form of thin films of $^{10}$B-enriched boron
carbide~\cite{b4cfilms_carina,Andersen2013116}. The basic principle of a
successful {\em direct detection} event in the latter is illustrated on the left
in fig.\ \ref{fig:boron_principle_with_backscattermat}: after conversion
$\text{n}+^{\,10}\text{B}\rightarrow\alpha+\text{Li}^7(+\gamma)$ one of the
released ions travels into the instrumented counting gas where it can be detected
like any energetic charged particle would. Typically the instrumentation would
include anode wires, with the boron-carbide itself acting as the cathode, and
such a detector therefore has the inherent high-rate capability of any such
gaseous detector in proportional mode. Furthermore, due to the higher amount of
energy released in the conversion reaction involving $^{10}$B (2.3 MeV) compared
to the corresponding $^3$He reaction (0.77 MeV), and the implied large signals,
the technology offers the possibility for very high suppression of gamma
backgrounds~\cite{khaplanov2013} which can otherwise be a problem at neutron
instruments. Additionally it is a relatively cheap technology, allowing for
large detector coverage when needed. However, the main limitation of the method
is that high conversion efficiencies ($>50$\%
several tens of micrometers of converter, whereas the resulting $\alpha$ and Li ions
only have a reach of a few micrometers in solid materials. Thus, to obtain higher
detection efficiencies, one will typically either place many independent layers
of gas-facing converters in the neutron path or try to arrange the geometry so
as to keep the angle of incidence of the neutron on the converter as high as
possible, or a combination of the two\footnote{In this paper we adopt the
  convention that a 0$^\circ$ incidence angle correspond to normal incidence,
  whereas 90$^\circ$ correspond to grazing incidence.}. Detailed analytical
calculations of detection efficiency depending on inclination and converter
thickness exist~\cite{piscitelli2013}.  However, such solutions come at a penalty
of increased complexity and cost, and the present paper instead investigates the
performance of an alternative approach in which detection efficiency is
increased by the addition of a strongly scattering material at the back of the
detector.

\section{Recovering unconverted neutrons through back-scattering}

\begin{figure}[t!]
  \begin{center}
    \includegraphics[width=0.99\columnwidth]{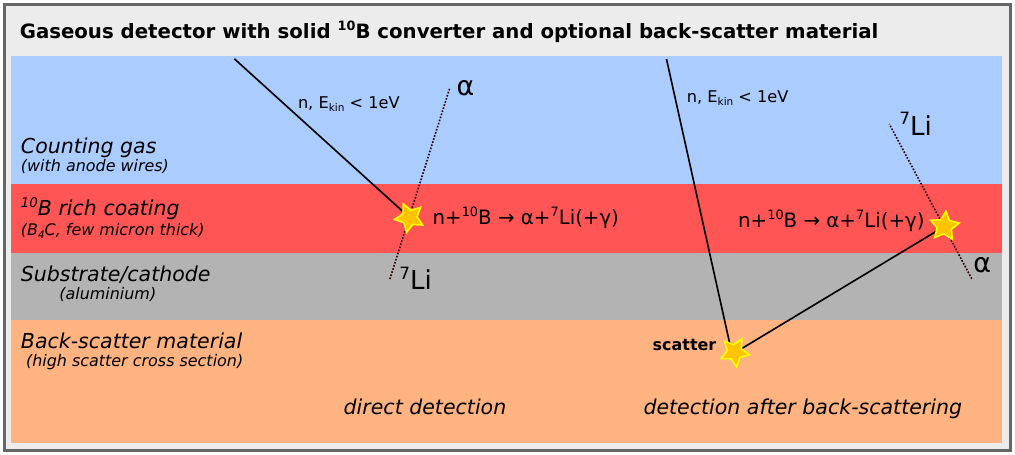}
    \caption{\label{fig:boron_principle_with_backscattermat}Principle of a $^{10}$B-based thin-film
      detector for thermal and cold neutrons.}
  \end{center}
\end{figure}

The principle of the method investigated in this paper is simple, as illustrated
on the right in fig.\ \ref{fig:boron_principle_with_backscattermat}: Placing a
material with high back-scattering cross-section behind a neutron detector will
increase detection efficiency, since a neutron which did not convert in the
detector in the first place will have a nonzero probability of scattering back
to the detector and converting the second time around, thus recovering events
which would otherwise go undetected.

Obviously, the downsides to adding such a scattering material would be expected
to be two-fold: Not only would the events thus recovered suffer from degraded
position resolution as well as a systematic positive shift in detection time,
but one would also potentially have to worry about back-scattered neutrons
escaping the detector through the front. The latter could potentially be a
problem if the instrument in question features detectors at opposite sides of a
sample in which neutrons scatter before detection. Collimators in front of the
detectors and shielding between detector sections would often be needed in any
case, and can be expected to significantly lower the negative impact of this
effect. Additionally, if the chosen back-scatter material primarily scatters
incoherently, one would avoid unduly adding new features to the detected
distributions. Naturally, for any given instrument one would need to carefully
analyse the exact implementational details of proposed designs in order to
ensure that all such effects would be understood, preferably through a
combination of tests on prototypes and simulations of the complete envisioned
setup.

The requirement of high cross-sections for incoherent scattering of thermal
neutrons suggests that a hydrogen-rich moderator-like material would be a good
candidate back-scatter material. From an operational and costing point of view,
the most suitable candidate would a priori seem to be a plastic such as the
widely used and studied moderator and shielding material, polyethylene (PE),
which is therefore the material we will focus on in the present investigations.

Regarding the placement of the back-scatter material, it should ideally be
immediately after, or as close as possible to, the last conversion material in
the detector. This, in order to reduce the additional time shift and spatial
displacement of the back-scattered neutrons before a potential
conversion. Furthermore, it is obviously important that any material between the
last conversion material and the back-scatter material will have a very low
probability of neutron absorption. Finally, it is also clear that multilayered
detectors in which the depth-of-interaction is used to estimate the total
time-of-flight and hence the energy of the neutron, would not be suitable to be
extended with a back-scatter material.

Given the above considerations, the present investigations will focus on the
performance impact of adding a layer of polyethylene behind a single-layered
$^{10}$B-thin film based gaseous detector, as is illustrated in fig.\
\ref{fig:boron_principle_with_backscattermat}. Although not strictly specified
by the method, the detector will for simplicity be taken to be implemented as in
the figure with a converter-coated aluminium substrate at the backside of the
gas only, and with the polyethylene located at the back of the
substrate. Obviously, a multilayer setup at an instrument which does not rely
on time-of-flight information would also be possible, but in general multilayer
setups already come with high detection efficiencies and price tags, and the
presently discussed method is likely more suitable for the opposite scenarios.

\section{Simulations}

The exact impact upon instrument and detector performance of adding a
back-scatter material to a detector at a neutron scattering instrument will
depend in detail on the detector technology and layout, the instrument purpose
and the exact integration of the back-scatter material itself into the
design. But it is nonetheless possible to evaluate the feasibility of the
back-scattering concept by considering a specific and simple, yet reasonably
realistic setup. The present investigations will therefore consist of realistic
Monte Carlo simulations of a setup exactly like the one depicted in fig.\
\ref{fig:boron_principle_with_backscattermat}, in which layers of counting gas,
converter, substrate and possibly a back-scatter material will be placed in the
path of incoming neutrons. The layers will be assumed to be very large in the
transverse dimensions.

\subsection{Setup details and validation}

\begin{figure}[t!]
  \begin{center}
    \includegraphics[width=0.99\columnwidth]{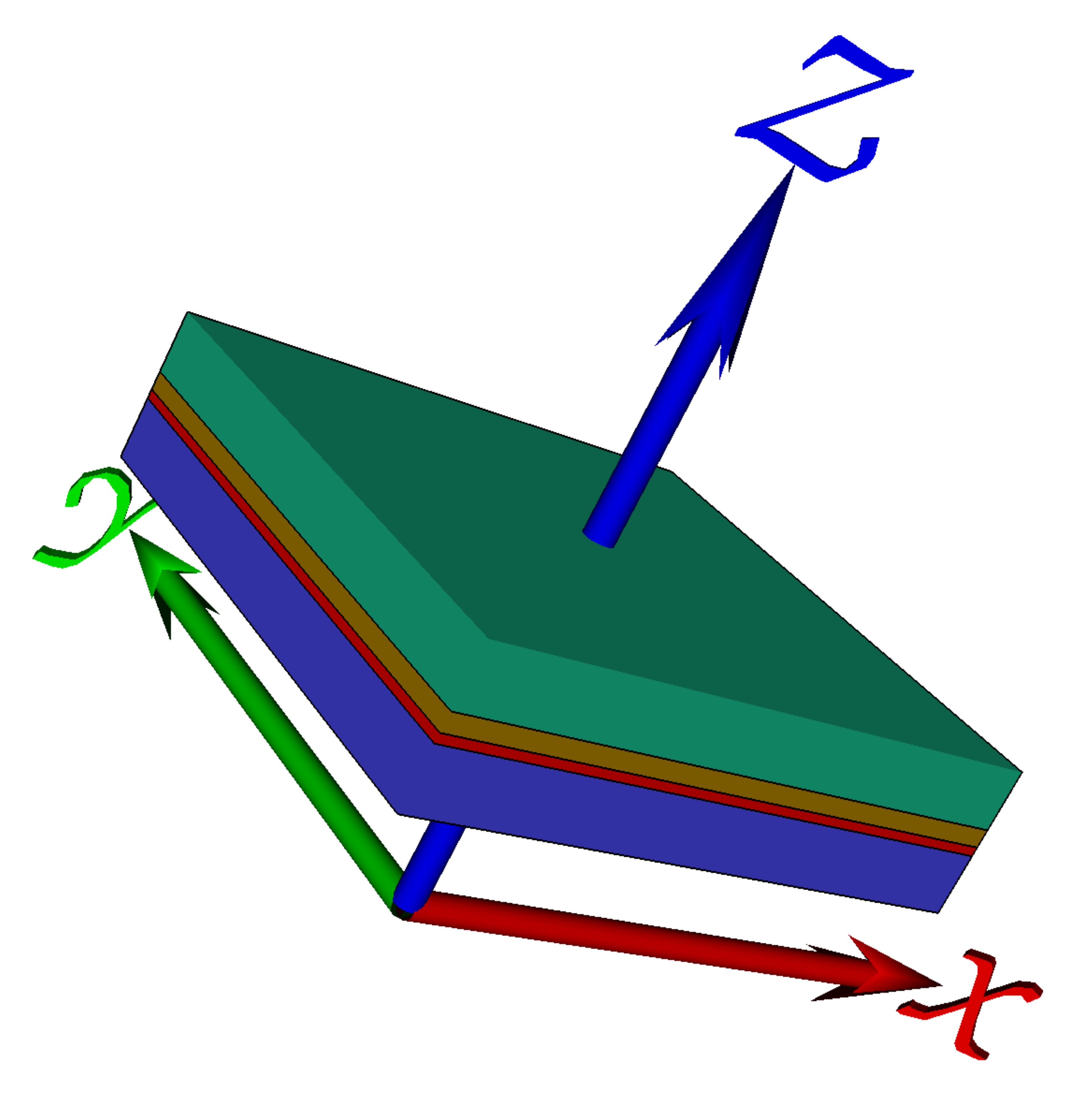}
    \caption{\label{fig:recovergeometry_geant4}Visualisation of the geometry
      used for the simulations: Neutrons are shot in the direction of the
      $z$-axis (blue arrow), which in the shown configuration is at normal
      incidence to the detector plane, and encounters planes of counting gas
      (blue), converter (red), substrate (brown) and scatter material (green).
      For clarity of visualisation purposes, the converter and substrate
      thicknesses have been blown up and the transverse extent of the plane
      reduced. }
  \end{center}
\end{figure}

The simulations are implemented and carried out using the
\texttt{Geant4}~\cite{geant4a,geant4b}-based framework described
in~\cite{dgcodechep2013}, and the simulated geometry is shown in fig.\
\ref{fig:recovergeometry_geant4}. For reference, the present investigations are
implemented in \texttt{Geant4} version 10.0.3 with a 70\%
counting gas mixture of Ar-CO$_2$, a boron-carbide converter enriched to 98\%
$^{10}$B, and gas and substrate plane thicknesses of 10mm and 0.5mm
respectively. The results are not believed to be sensitive to any of these
choices.

Previous studies~\cite{macrostruct2013}, which also included comparisons with
test-beam data, have shown that \texttt{Geant4} with the \texttt{QGSP\_BIC\_HP}
physics list is able to adequately capture the physics of the neutron absorption
in the enriched boron-carbide and the subsequent journey of the released ions
into the counting gas. It was also shown that a simple threshold, here chosen to
be 150keV, on the amount of energy deposited therein suffices to accurately
emulate actual detection efficiencies in a real detector equipped with anode
wire-planes in the counting gas. This is of course not surprising, as one would
first of all expect \texttt{Geant4} to describe absorption of thermal and colder
neutrons well, given that the relevant cross-sections are given by a simple
$\sim1/v_\text{n}$ law. Secondly, because a good description of energy loss by
charged particles in matter is an essential feature of \texttt{Geant4}, used as
it is to routinely model energy depositions in a broad range of applications.

\begin{figure}[t!]
  \begin{center}
    \includegraphics[width=0.99\columnwidth]{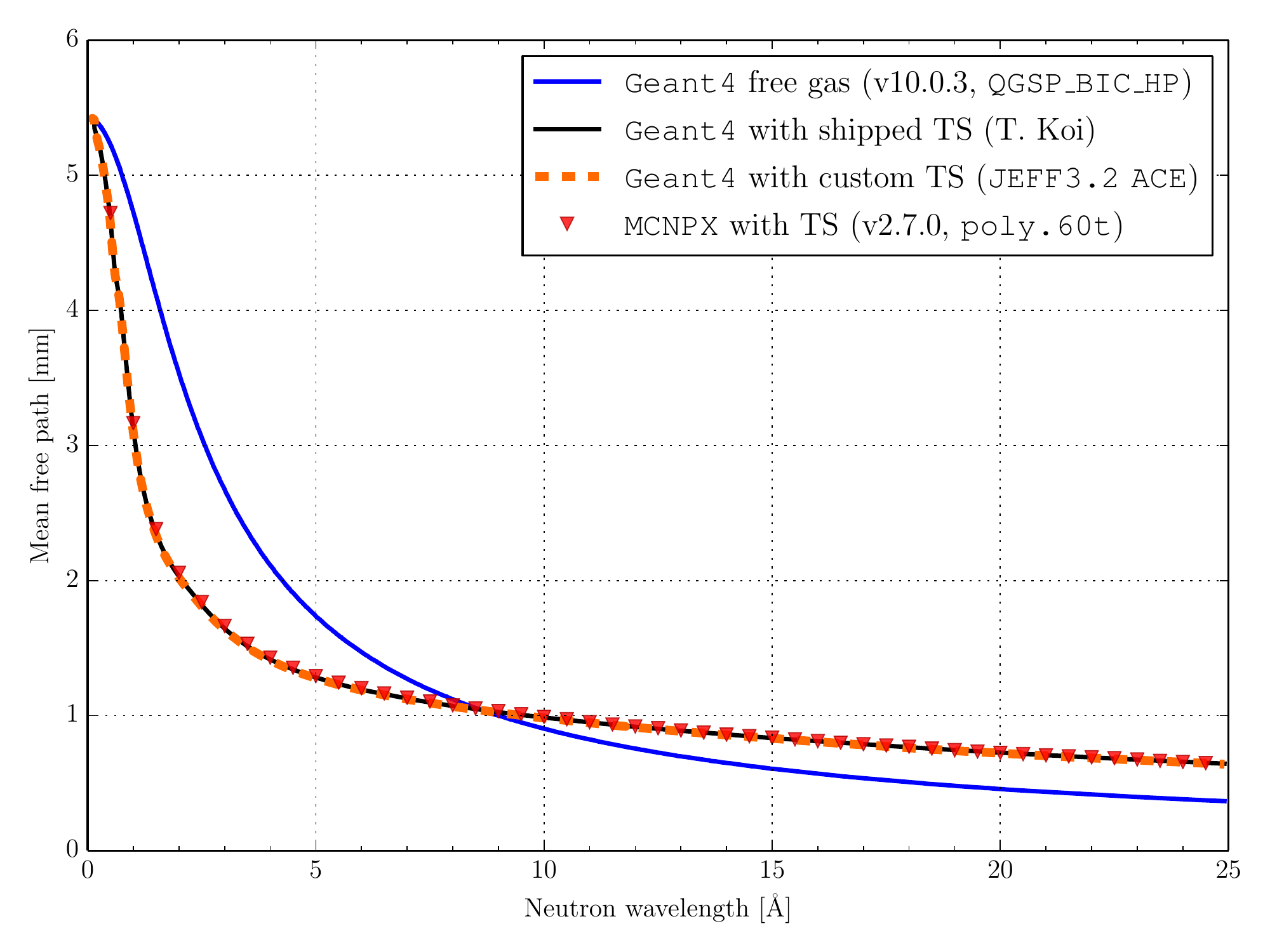}
    \caption{\label{fig:g4_pe_xsects} Mean free path of neutrons in polyethylene
      in \texttt{Geant4} and \texttt{MCNPX}, as a function of neutron wavelength
      and for different physics models. The methods for extracting these values
      are discussed in the text.}
  \end{center}
\end{figure}

However, materials in \texttt{Geant4} are usually described under a free-gas
assumption, with no information concerning inter-atomic chemical bindings. This
means that potentially important features like Bragg-diffraction in a
polycrystalline material like aluminium, or thermal scattering (TS) on energy
levels in hydrogen bonding in polyethylene are a priori missing. For the former
issue, the setup is therefore augmented to use the \texttt{NXSG4}
extension~\cite{nxsg4}, to ensure accurate description of the aluminium used in
the substrate. For the latter issue, it is fortunate that polyethylene, seeing
significant use for neutron moderation and shielding, is one of a select few
materials for which detailed thermal scattering cross-section data at various
temperatures exists, although different codes and data versions might provide
slightly different results. Here, we will compare a model shipped with
\texttt{Geant4} itself, due to work by T.\ Koi, with a custom in-house model to
evaluate the JEFF-3.2 ACE formatted files for the Thermal Scattering
Law~\cite{jeff2011}, as well as, for reference, with output from simulations
carried out with \texttt{MCNPX} version 2.7.0~\cite{mcnpx2011} using the
\texttt{poly.60t} thermal cross sections derived from ENDF6.5 dynamic structure
factors using NJOY~\cite{njoy2012}. Choosing pragmatically a material
temperature of 293.6K which is available for polyethylene in all three
implementations, and comparing also with the base free-gas treatment in
\texttt{QGSP\_BIC\_HP}, the resulting mean free path lengths of neutron
interactions in polyethylene are shown in fig.\ \ref{fig:g4_pe_xsects}. Note
that for reliability, the values were extracted at run-time rather than
second-guessed from data files. For \texttt{Geant4}, this was done via a custom
hook querying the physics processes \texttt{GetMeanFreePath(..)} methods. For
\texttt{MCNPX}, the values were determined through direct simulations of
neutrons impinging on a very thin plane of polyethylene, and the resulting
statistical errors are smaller than the plot markers. The importance for our
analysis of using specific thermal scattering data for polyethylene, rather than
just the base free-gas model is clear, as the resulting mean free path length
for neutrons scattering in polyethylene is affected with as much as a factor of
2.

\begin{figure}[t!]
  \begin{center}
    \includegraphics[width=0.99\columnwidth]{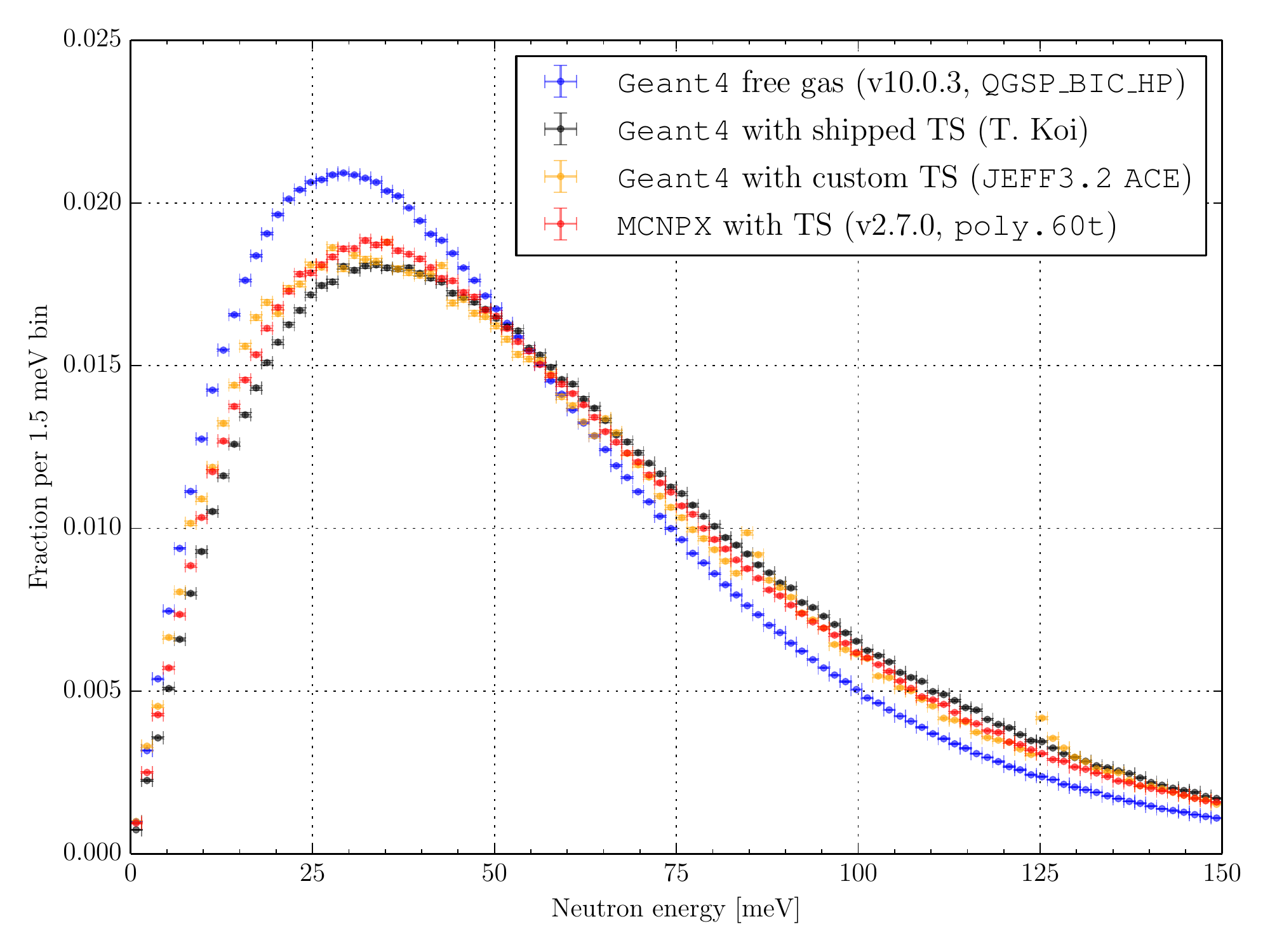}
    \caption{\label{fig:ts_50mm_pe_transmission}Simulated transmission spectrum
      for 2.5\AA\ neutrons through 50mm polyethylene in both \texttt{Geant4}
      with different physics models and \texttt{MCNPX}.}
  \end{center}
\end{figure}

\begin{figure}[t!]
  \begin{center}
    \includegraphics[width=0.99\columnwidth]{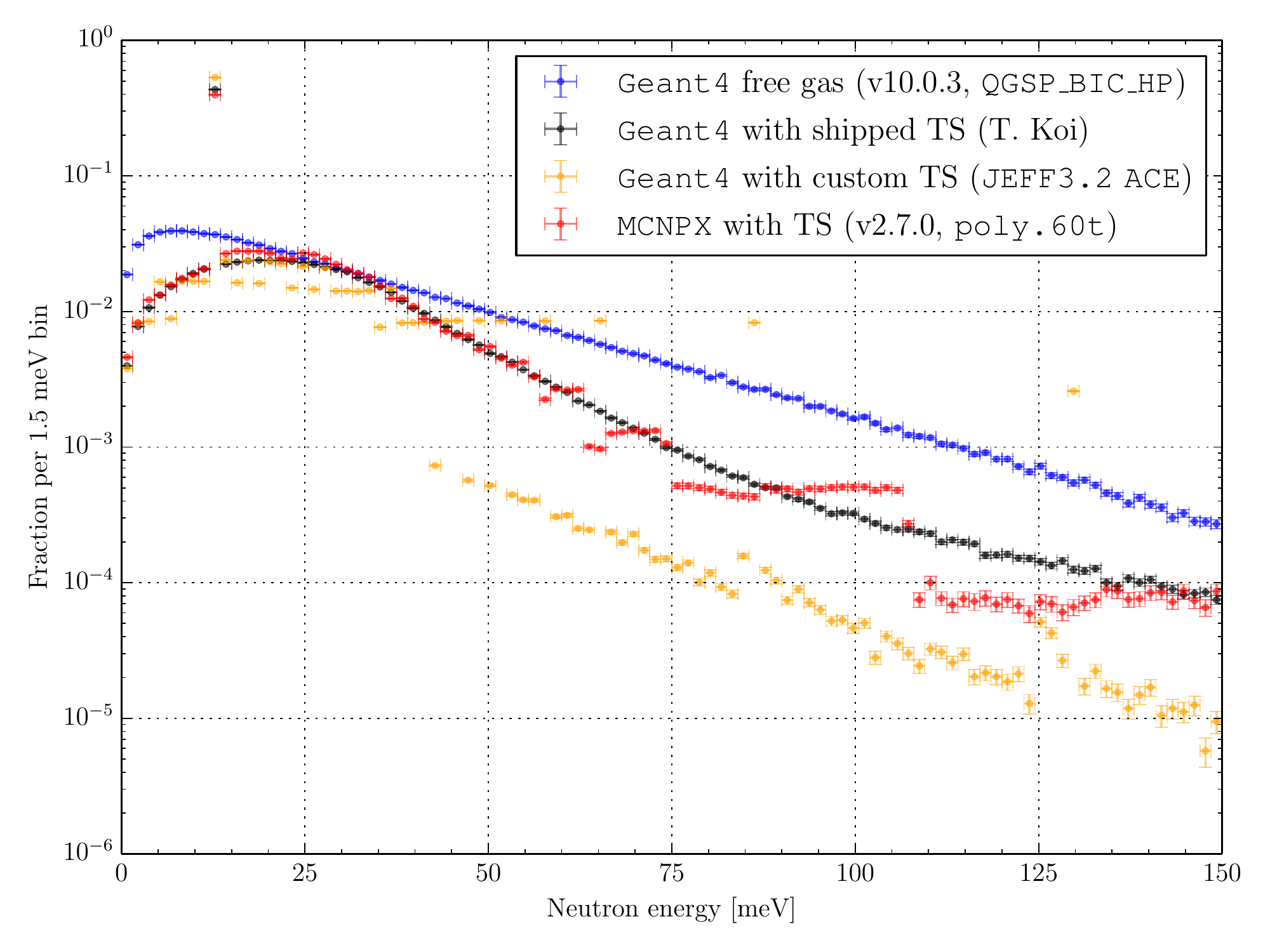}
    \caption{\label{fig:ts_100micron_pe_backscatter} Simulated back-scatter
      spectrum for 2.5\AA\ neutrons on 100$\mu$m polyethylene in both
      \texttt{Geant4} with different physics models and \texttt{MCNPX}.}
  \end{center}
\end{figure}

As a further comparison, fig.\ \ref{fig:ts_50mm_pe_transmission} shows the
simulated transmission spectrum of 2.5\AA\ neutrons moderated in a 50mm thick
slab of polyethylene. For the purposes of the present investigations, the three
curves with specific thermal scattering models are compatible and different from
the free-gas model. However, as the present investigations involve
back-scattering in potentially very thin layers of polyethylene, it is
interesting to compare the back-scattered spectrum after a single
interaction. Thus, fig.\ \ref{fig:ts_100micron_pe_backscatter} shows a
comparison of the back-scattered spectrum in a 100$\mu$m polyethylene slab, of
which two interesting observations can be made: First of all, the three thermal
scattering models all include an incoherent elastic peak at the energy of the
incoming neutron (2.5\AA\ $\sim$ 13.1meV), whereas this is absent in the free-gas
model. Secondly, whereas it is once again clear that the general shapes of
the distributions for the three thermal scattering models follow the same rough
shape, it is only the one shipped with \texttt{Geant4} which does not exhibit
artifacts due to the discrete parameterisations of the thermal scattering data,
and for that reason this is the model chosen for the investigations in the
following sections.

\subsection{Results}

\begin{figure}[t!]
  \begin{center}
    \includegraphics[width=0.99\columnwidth]{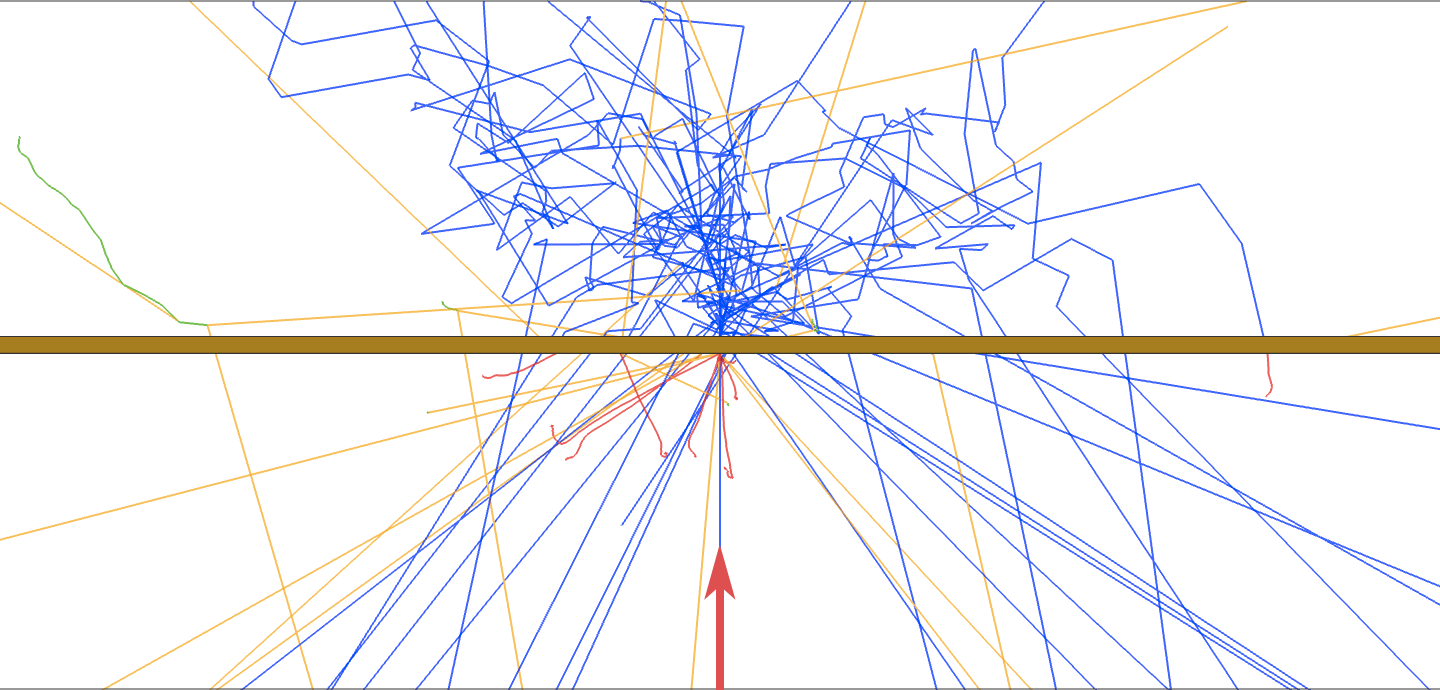}
    \caption{\label{fig:sim_example}Simulated trajectories of 50
      1.8\AA\ neutrons (blue) starting in the center of the bottom edge of the
      figure (indicated by a red arrow) and travelling upwards through layers of
      counting gas, 2.5$\mu$m of enriched boron-carbide, 0.5mm of aluminium
      substrate (brown) and finally 10mm of polyethylene. As a result of neutron
      absorption, secondary particles are released in various locations: photons
      (orange), electrons (green) and $\alpha$ or $\text{Li}^7$ ions (red).}
  \end{center}
\end{figure}

As an illustration, simulated trajectories of particles resulting from firing 50
neutrons into our setup are visualised in fig.\ \ref{fig:sim_example}: neutrons
arriving from below travel through the counting gas and either convert in the
enriched boron-carbide immediately, releasing charged ions into the gas, or pass
onwards through the substrate and into the polyethylene where complex
trajectories follow as a result of the multiple scattering
interactions. Ultimately those neutrons either pass through to the backside of
the polyethylene, get absorbed in processes releasing photons, or return back to
the substrate and converter, where some are absorbed the second time around as
hoped, again releasing ions into the counting gas.

The key issues to be investigated with the simulations are on one hand what
positive impact placing different amounts of polyethylene will have on detection
efficiency, and on the other hand what the associated negative impacts on time
and position resolution will be, and to what extent those can be mitigated. The
answers can be expected to depend also on the angle of incidence of the neutron
on the detection plane, the neutron energy and on the chosen thickness of the
converter coating. To answer these questions, large numbers of neutrons were
simulated for a variety of configurations, and the results subsequently
carefully analysed in order to produce the plots in this section. For reference,
a total of $3.5\cdot10^{10}$ neutron events were simulated, using between
$4.0\cdot10^6$ and $2.0\cdot10^8$ for each given configuration, as
needed to achieve reasonably low statistical fluctuations in all plots.

\begin{figure}[t!]
  \begin{center}
    \includegraphics[width=0.99\columnwidth]{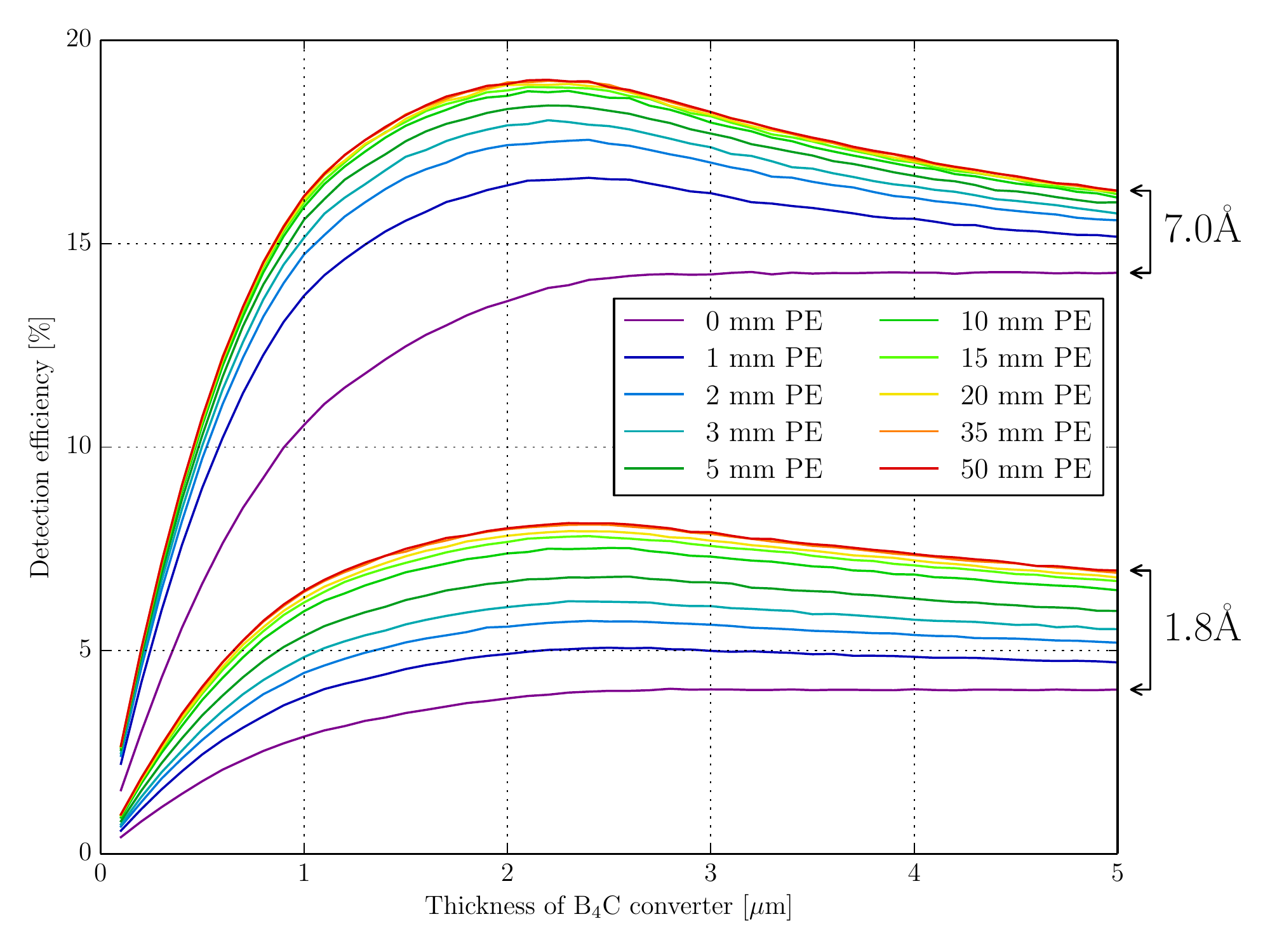}
    \caption{\label{fig:converter_thickness_effect_0deg} Simulated detection
      efficiency for neutrons at normal incidence as a function of converter
      thickness, for various neutron energies and amounts of polyethylene.}
  \end{center}
\end{figure}

First, simulated detection efficiencies at normal incidence as a function of
converter thickness are shown in fig.\
\ref{fig:converter_thickness_effect_0deg} for various amounts of polyethylene
and for both thermal (1.8\AA) and cold (7.0\AA) incident neutron energies. As
expected, detection efficiencies for colder neutrons are higher, but otherwise
the qualitative features of the curves are similar to those at thermal
energies. For simplicity, the remainder of our investigations will therefore
focus on neutrons with a wavelength of 1.8\AA. The next thing to notice is that
the curves without polyethylene grow monotonically with converter thickness, but
effectively saturate at the final plateau already around $2.5\mu$m, which is the
effective range in the converter of the ions released in the conversion.  These
converter thicknesses compare well to the expectation from the analytical
calculations~\cite{piscitelli2013}, lending further credence to the simulation.
On the other hand, when polyethylene is added, the curves exhibit a maximum
around 2.1--$2.5\mu$m, which is expected since conversions at ever deeper
locations are increasingly unlikely to contribute positively with direct
detection events, and will merely act as unwanted inactive shielding in the path
of neutrons being back-scattered by the polyethylene. Fortunately, it appears
that all curves are reasonably close to their maximum value when the converter
thickness is set to $2.5\mu$m, allowing for the simplifying assumption of using
this value throughout the remainder of the investigations, without unduly
biasing the comparisons. Finally, it appears as could be expected that detection
efficiencies only grow with the amount of polyethylene added, but at an ever
decreasing rate --- the biggest gain coming from the first few millimetres added.

\begin{figure}[t!]
  \begin{center}
    \includegraphics[width=0.99\columnwidth]{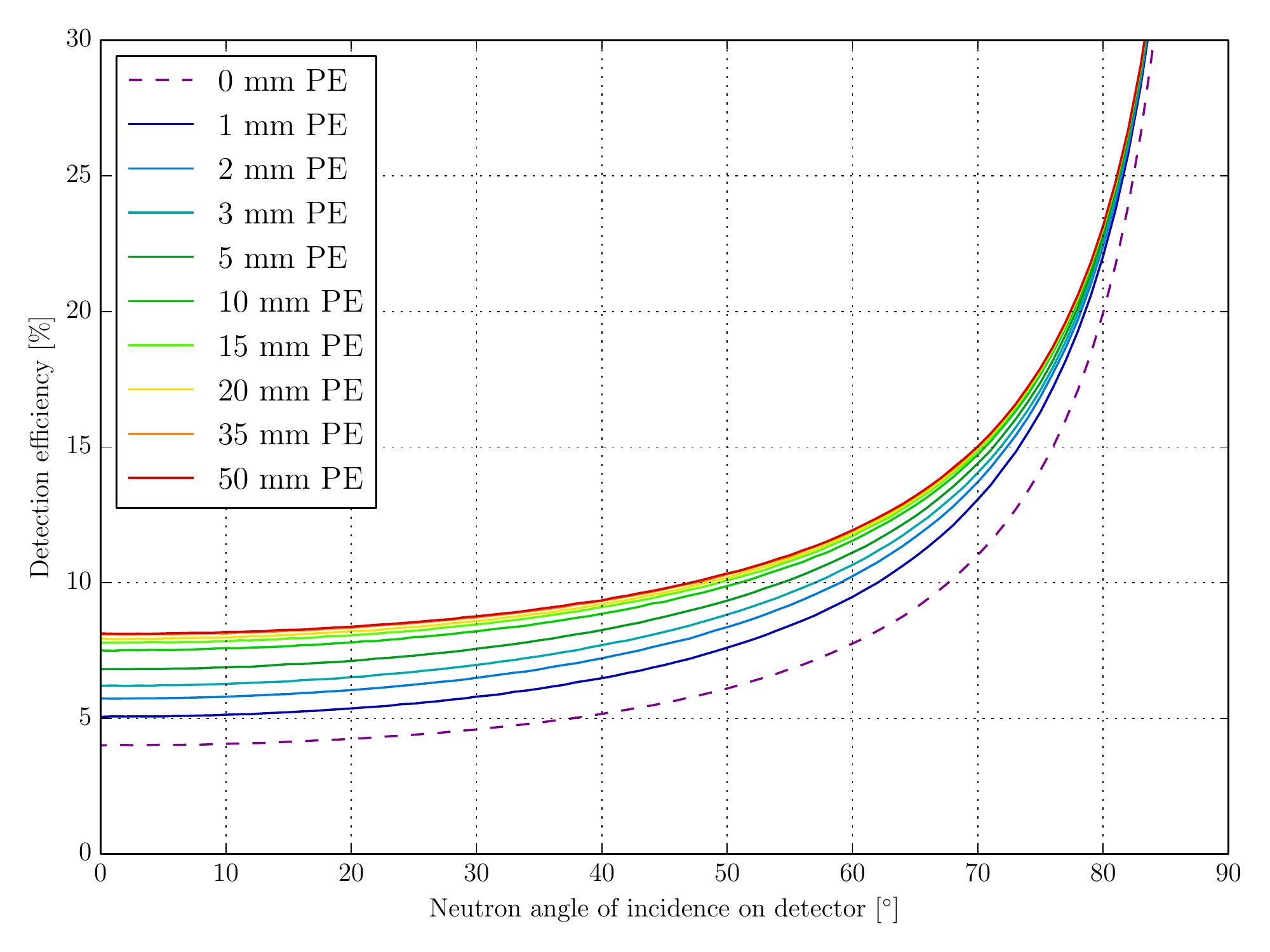}
    \caption{\label{fig:pe_effect} Simulated detection efficiency for various
      amounts of polyethylene as a function of the incidence angle between the
      incoming neutron and the detector plane.}
  \end{center}
\end{figure}

\begin{figure}[t!]
  \begin{center}
    \includegraphics[width=0.99\columnwidth]{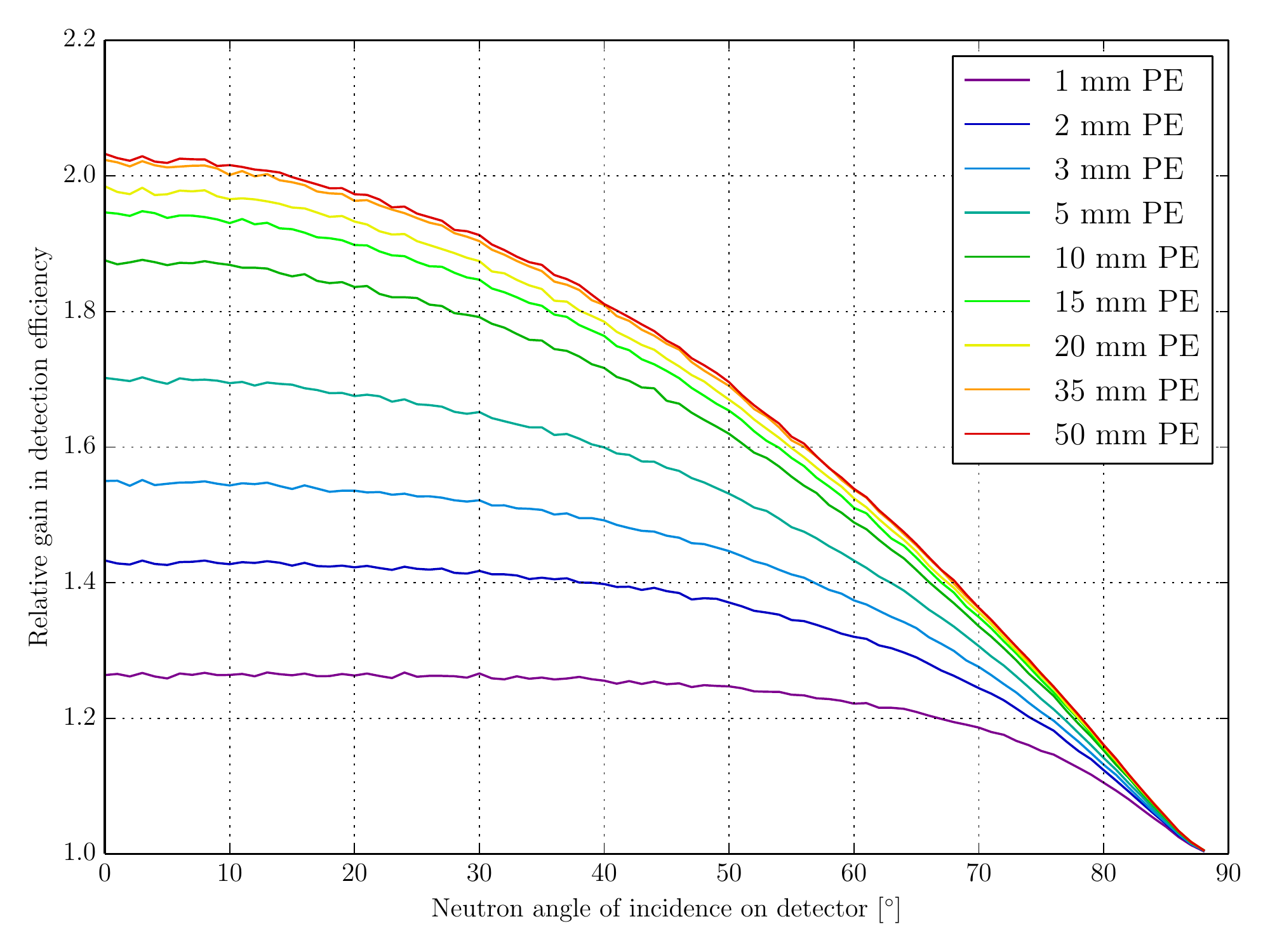}
    \caption{\label{fig:pe_effect_versus_ref} Simulated relative gain in
      detection efficiency for various amounts of polyethylene as a function of
      the incidence angle between the incoming neutron and the detector plane.}
  \end{center}
\end{figure}

Next, fig.\ \ref{fig:pe_effect} shows the detection efficiency as a function of
incidence angle of the neutron on the detector planes. Naturally, the efficiency
curves all increase sharply as the incoming neutron tend towards grazing
incidence. However, as is seen more clearly in fig.\
\ref{fig:pe_effect_versus_ref}, the relative gain in detection efficiency from
adding the polyethylene decreases at higher incidence angles. This is easily
understood since the back-scattering from the polyethylene is incoherent and
thus effectively isotropic: at low incidence angles a random back-scattered
neutron is likely to hit the converter at a higher incidence than during the
initial traversal, whereas at high incidence angles the situation is
reversed. Thus, while the result of fig.\ \ref{fig:pe_effect_versus_ref} is
promising very tangible improvements at low incidence angles, it is also making
it clear that the concept is not suitable for detectors which are to be operated
at higher incidence angles. For that reason, the rest of the present
investigations will focus on neutrons with low incidence angle (perpendicular to
the surface).

\begin{figure}[t!]
  \begin{center}
    \includegraphics[width=0.99\columnwidth]{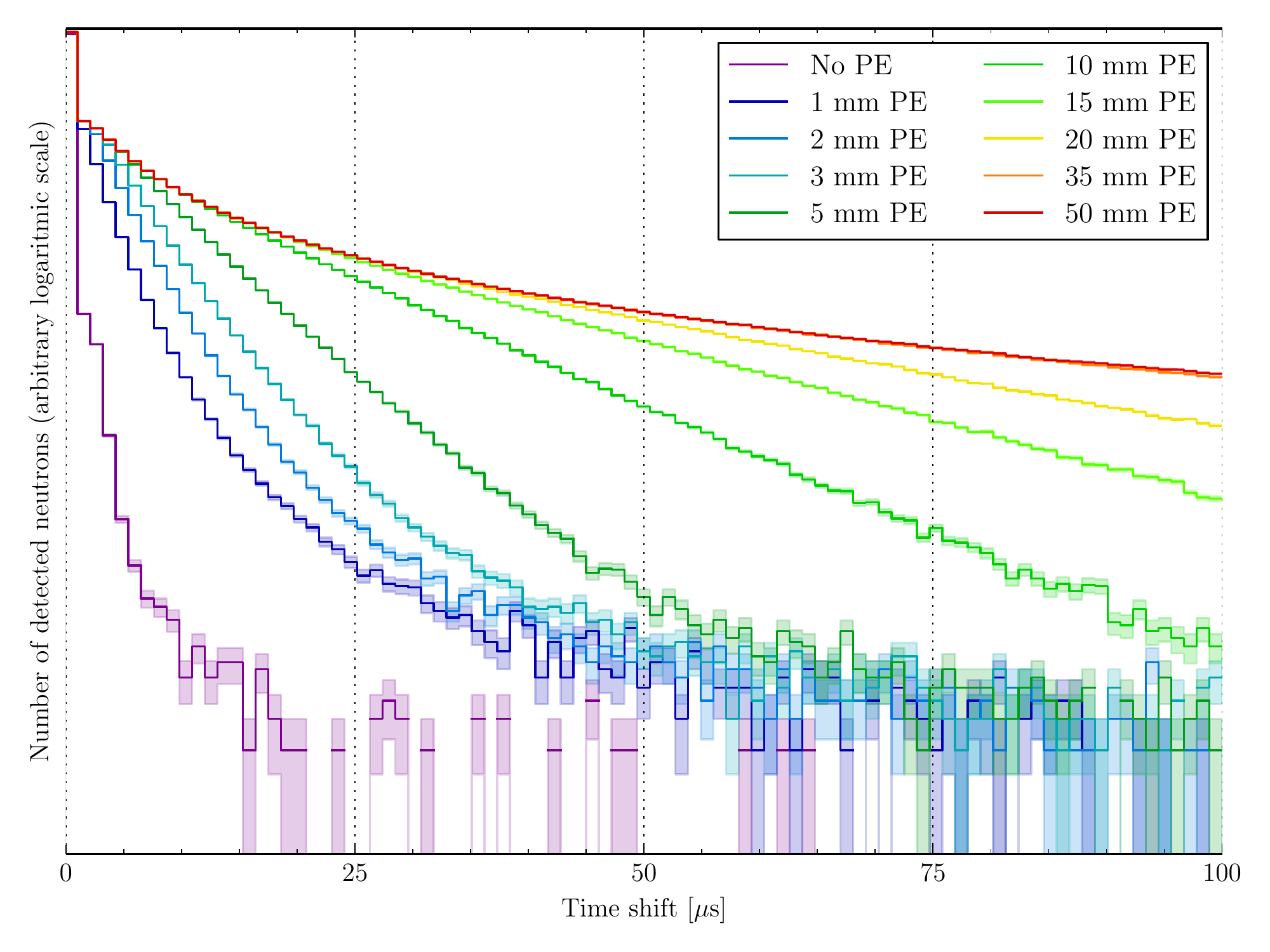}
    \caption{\label{fig:timespread_given_pe_thickness_abs} Simulated
      distribution of shift in detection time for various amounts of
      polyethylene.}
  \end{center}
\end{figure}

\begin{figure}[t!]
  \begin{center}
    \includegraphics[width=0.99\columnwidth]{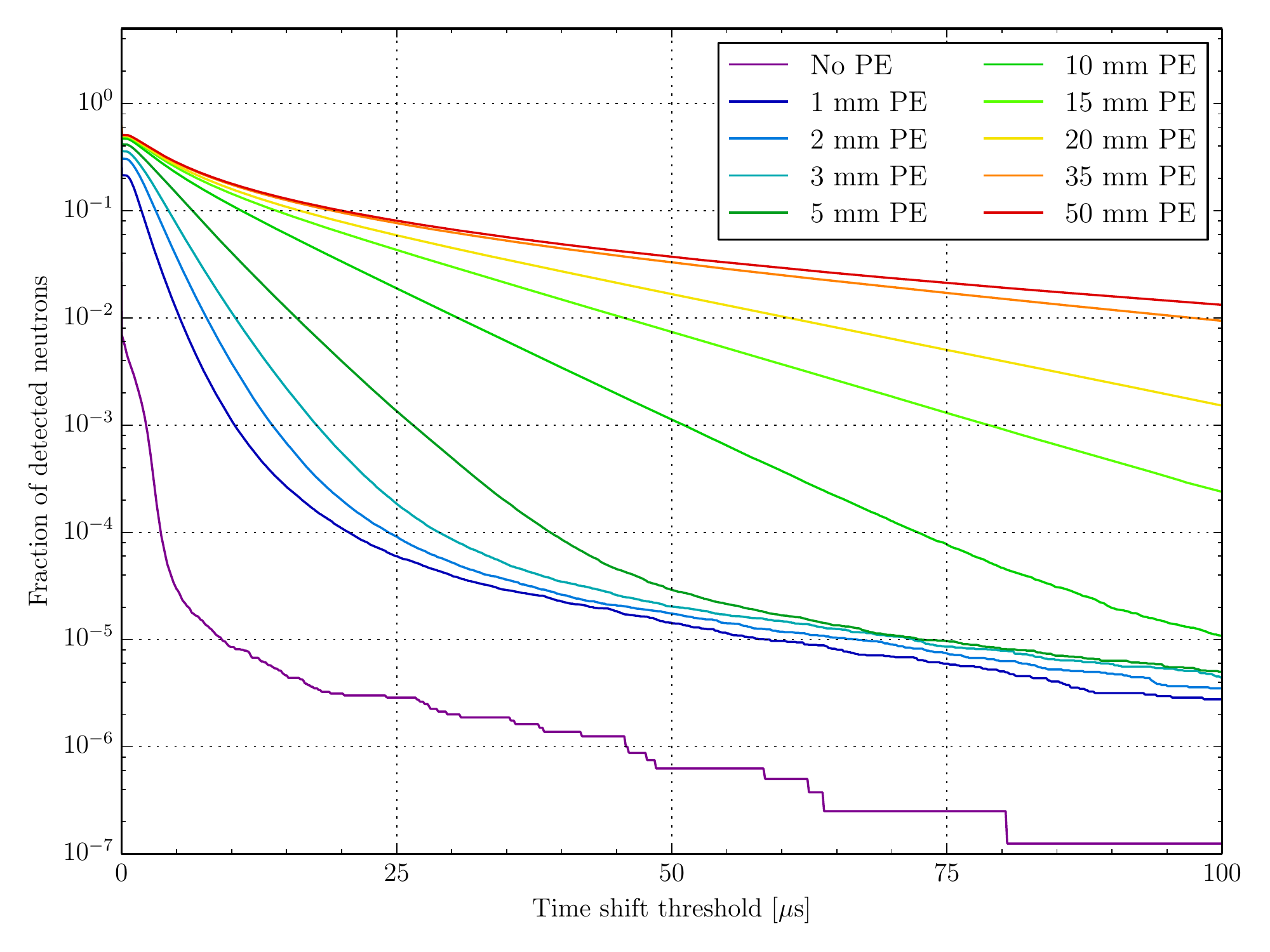}
    \caption{\label{fig:timespread_given_pe_thickness_commul} Simulated fraction
      of neutrons above a given shift in detection time for various amounts of
      polyethylene.}
  \end{center}
\end{figure}

Turning to the possible downsides, fig.\
\ref{fig:timespread_given_pe_thickness_abs} shows the distribution of simulated
detection time for various amounts of polyethylene, counting from the moment the
neutron first enters the converter and until the time when energy exceeding the
threshold is deposited in the countring gas, usually by the released ions. It is
clear that addition of polyethylene behind the detector leads to increasing
tails towards larger times, but in order to quantify the effects it is arguably
more useful to look instead at the curves showing the fraction of neutrons with
detection time shifts above a given threshold shown in fig.\
\ref{fig:timespread_given_pe_thickness_commul}. For instance, one can learn that
if a specific detector has a requirement that all except $10^{-3}$ of the
detected neutrons must be detected with a time shift less than 50$\mu$s, one
should not add more than $\sim$10mm of back-scattering polyethylene. Fortunately,
even a 100$\mu$s resolution at the $10^{-2}$ level would be adequate for many
neutron instruments~\cite{kirstein2014}.

\begin{figure}[t!]
  \begin{center}
    \includegraphics[width=0.99\columnwidth]{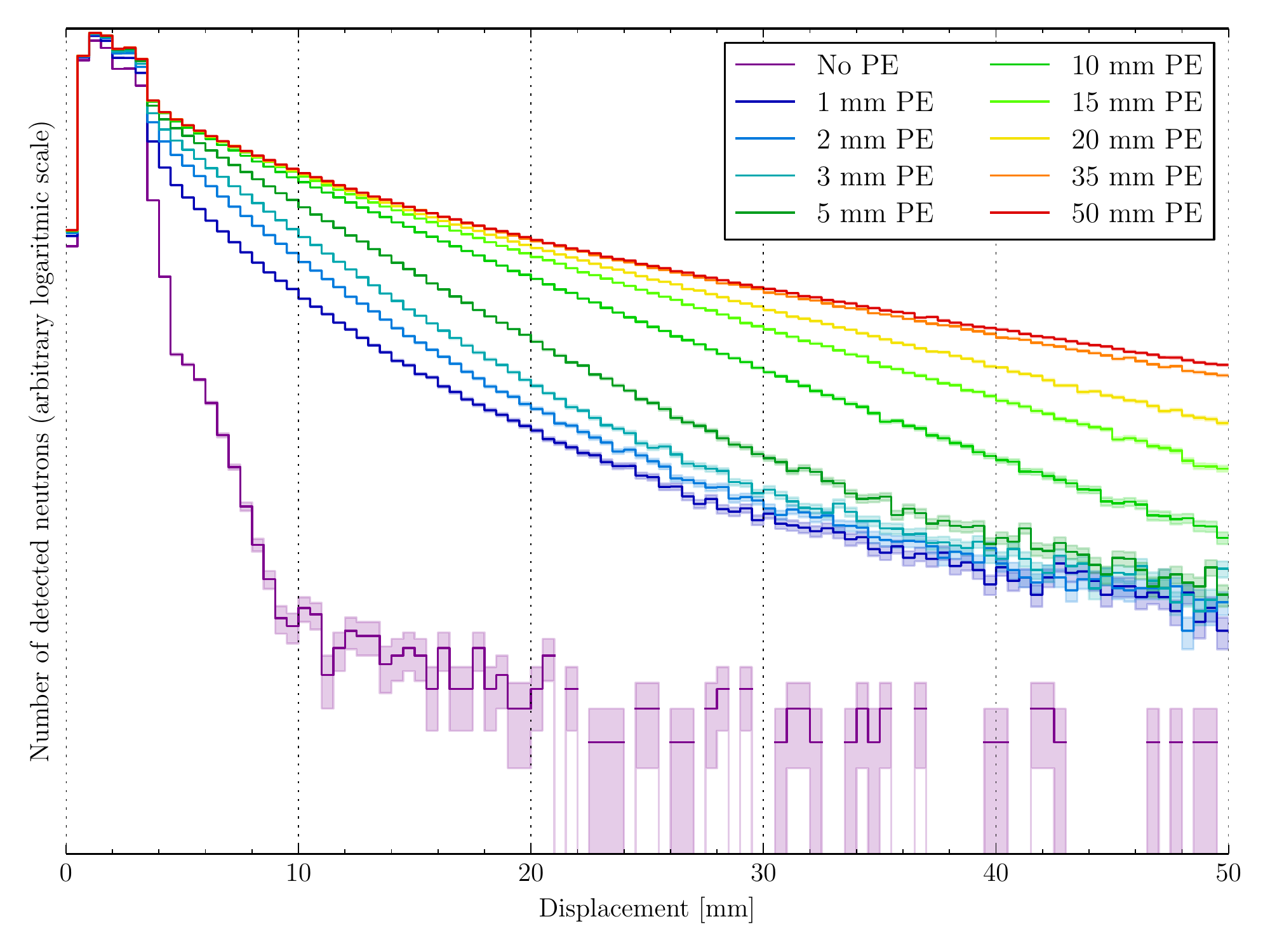}
    \caption{\label{fig:displacement_given_pe_thickness_abs} Simulated
      distribution of displacement in detection location for various amounts of
      polyethylene.}
  \end{center}
\end{figure}

\begin{figure}[t!]
  \begin{center}
    \includegraphics[width=0.99\columnwidth]{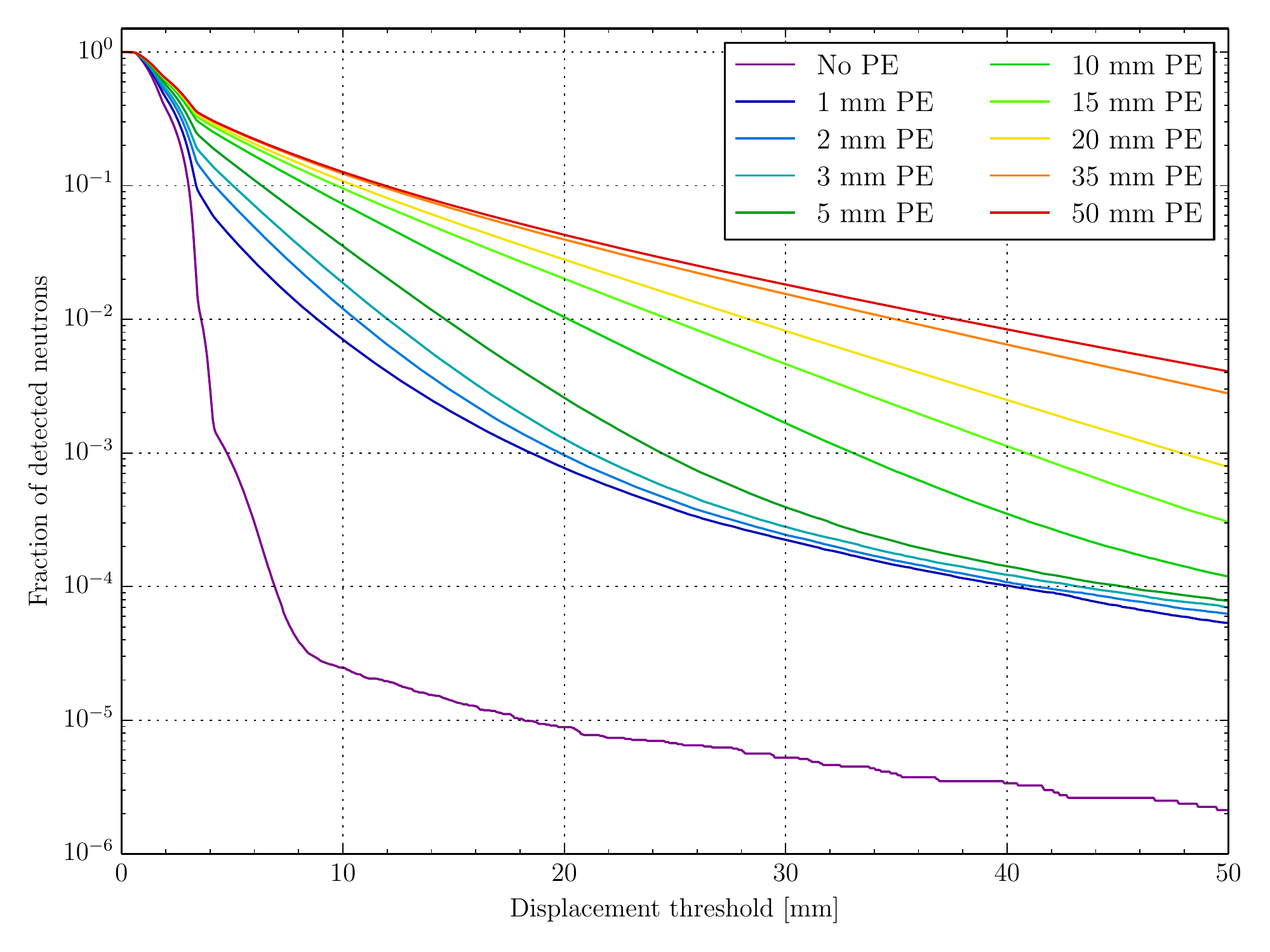}
    \caption{\label{fig:displacement_given_pe_thickness_commul} Simulated
      fraction of neutrons having a displacement in detection location above a
      given threshold, for various amounts of polyethylene.}
  \end{center}
\end{figure}

Next, fig.\ \ref{fig:displacement_given_pe_thickness_abs} and fig.\
\ref{fig:displacement_given_pe_thickness_commul} show the corresponding
distributions of the spatial displacement of the detection location, given by
the location of energy depositions in the counting gas, relative to the position
where the neutron first enters the converter. Again we can readily read of
performance metrics from the second of the figures: If one requires all except
$10^{-2}$ of neutrons to be detected with a displacement less than 20mm, no more
than 10mm of polyethylene should be added behind the detector.

\subsection{Possible mitigation strategies}

The results so far, summarised in fig. \ref{fig:pe_effect_versus_ref},
\ref{fig:timespread_given_pe_thickness_commul} and
\ref{fig:displacement_given_pe_thickness_commul}, indicate that the presented
method has the potential to provide substantial gains in detection efficiencies,
but with potentially significant associated adverse effects on both temporal and
spatial detection resolutions. In this section we will briefly investigate two
strategies for reducing the impact of these adverse effects. In essence, both
will try to eliminate some or all unfavourable trajectories in the polyethylene,
while hopefully retaining a large fraction of the beneficial ones. Ignoring
events with neutrons either absorbed inside or transmitted through (to an
absorbing backside most likely) the polyethylene, favourable trajectories are
obviously those where the neutron is back-scattered out of the front of the
polyethylene with as small a distortion of distance and time as
possible. Unfavourable trajectories are on the other hand those where the
neutron either spends a long time inside the polyethylene (impacting temporal
detection), or travels a long transversal distance inside the polyethylene
(impacting spatial detection).

\begin{figure}[t!]
  \begin{center}
    \includegraphics[width=0.99\columnwidth]{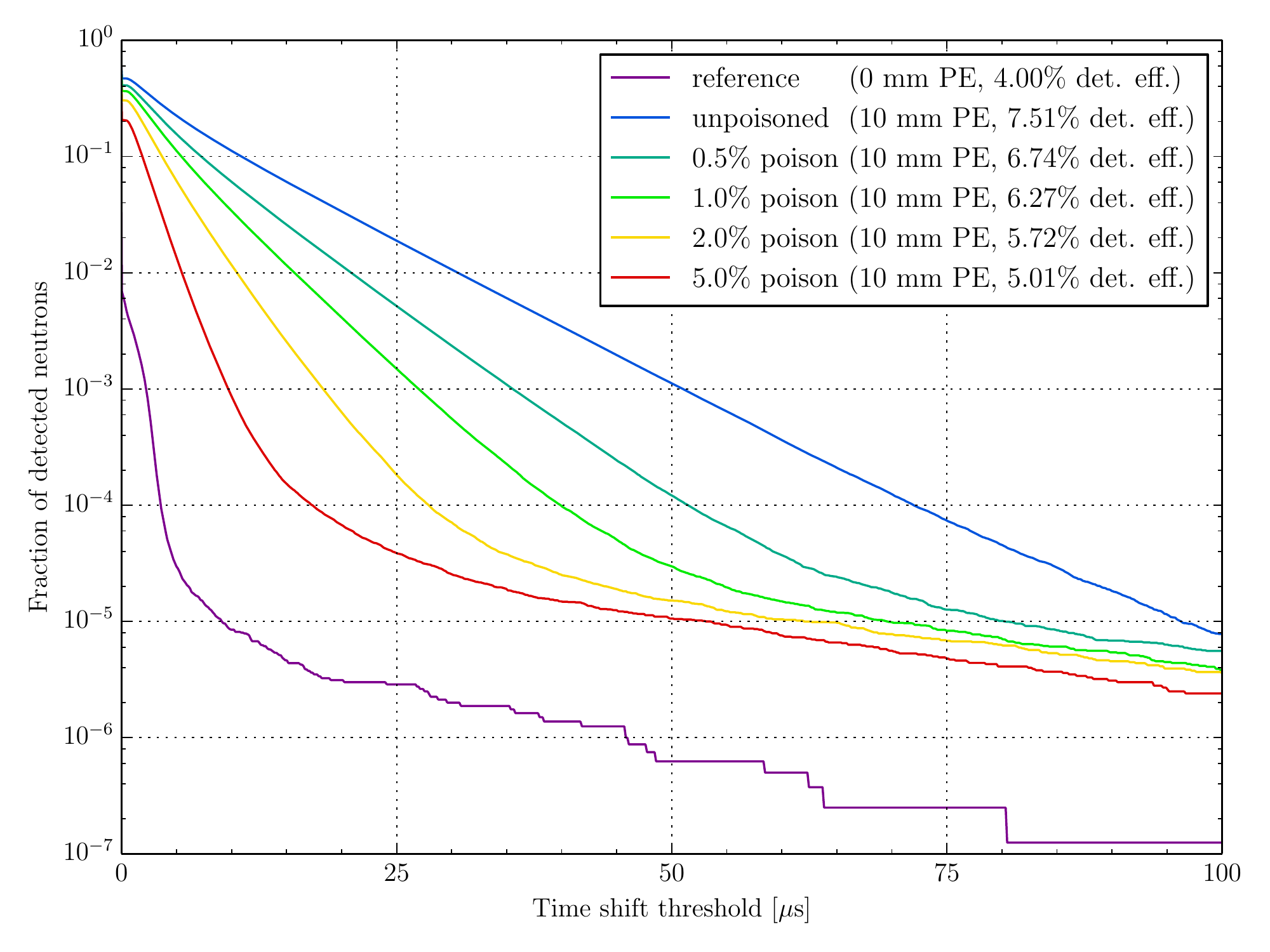}
    \caption{\label{fig:timespread_given_poison_commul} Simulated fraction of
      neutrons above a given shift in detection time for no polyethylene as well
      as 10mm polyethylene with various levels of poisoning.}
  \end{center}
\end{figure}

\begin{figure}[t!]
  \begin{center}
    \includegraphics[width=0.99\columnwidth]{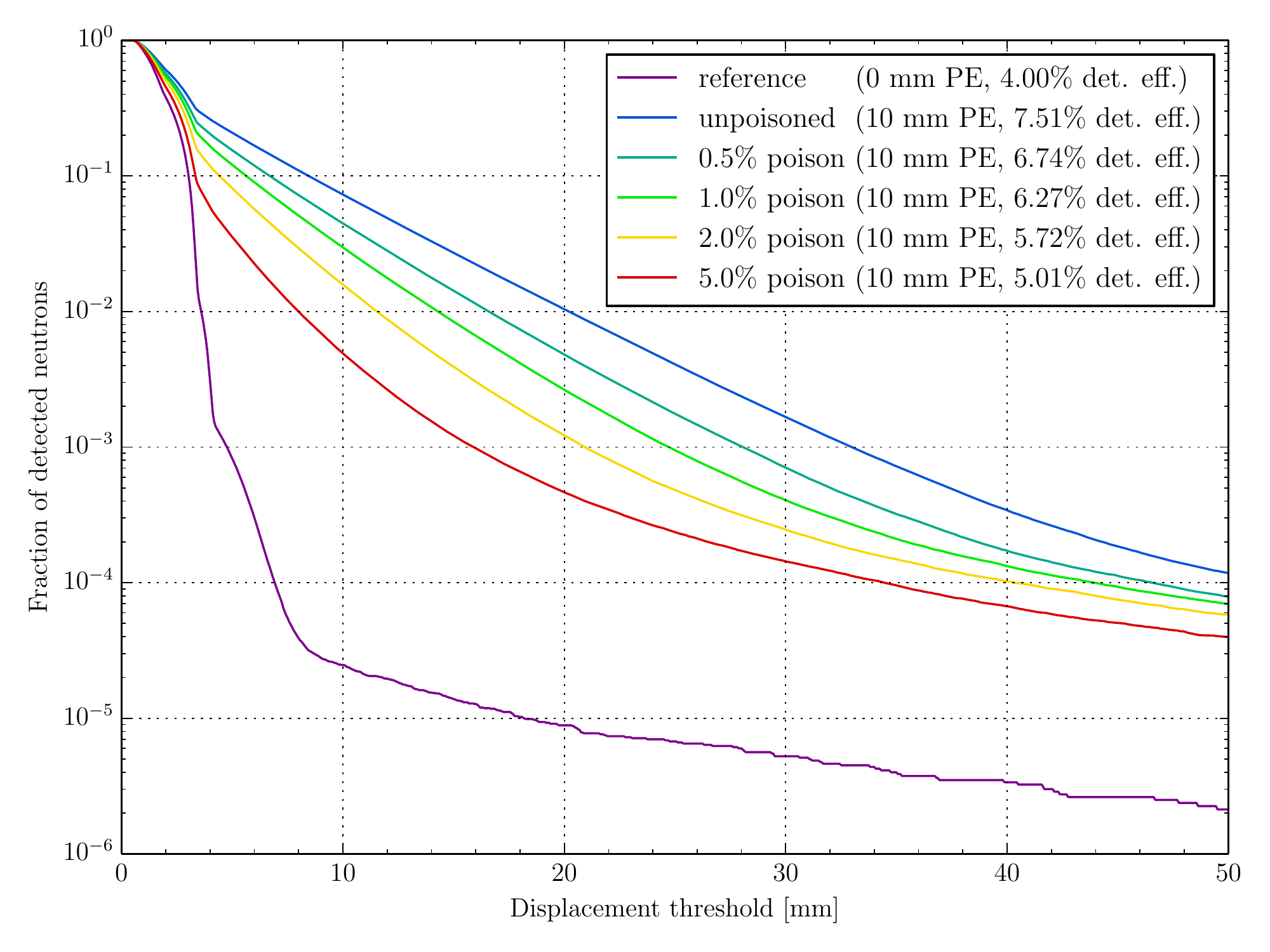}
    \caption{\label{fig:displacement_given_poison_commul} Simulated fraction of
      neutrons having a displacement in detection location above a given
      threshold, for no polyethylene as well as 10mm polyethylene with various
      levels of poisoning.}
  \end{center}
\end{figure}

To eliminate the first type of unfavourable trajectories, one might consider
``poisoning'' the polyethylene, by contaminating it with a small fraction of
atoms with high cross-section for neutron absorption. Done right, this should
ideally result in a large fraction of those neutrons spending a long time inside
the polyethylene being absorbed, with only little impact on neutrons promptly
back-scattered. To quantify the potential of poisoning, fig.\
\ref{fig:timespread_given_poison_commul} shows the simulated effect in a setup
with 10mm polyethylene. For example, adding 0.5\%
reduces the fraction of neutrons detected above 50$\mu$s by an order of
magnitude, while only reducing the detection efficiency from 7.51\%
the reference 4\%
potentially very potent method, if one is concerned with the detection time
resolution. On the other hand, fig.\ \ref{fig:displacement_given_poison_commul}
shows that the impact on the spatial resolution is somewhat smaller, which is as
expected since the unfavourable trajectories travelling a long transversal
distance in the polyethylene are not necessarily always spending a very long
time in it.

\begin{figure}[t!]
  \begin{center}
    \includegraphics[width=0.99\columnwidth]{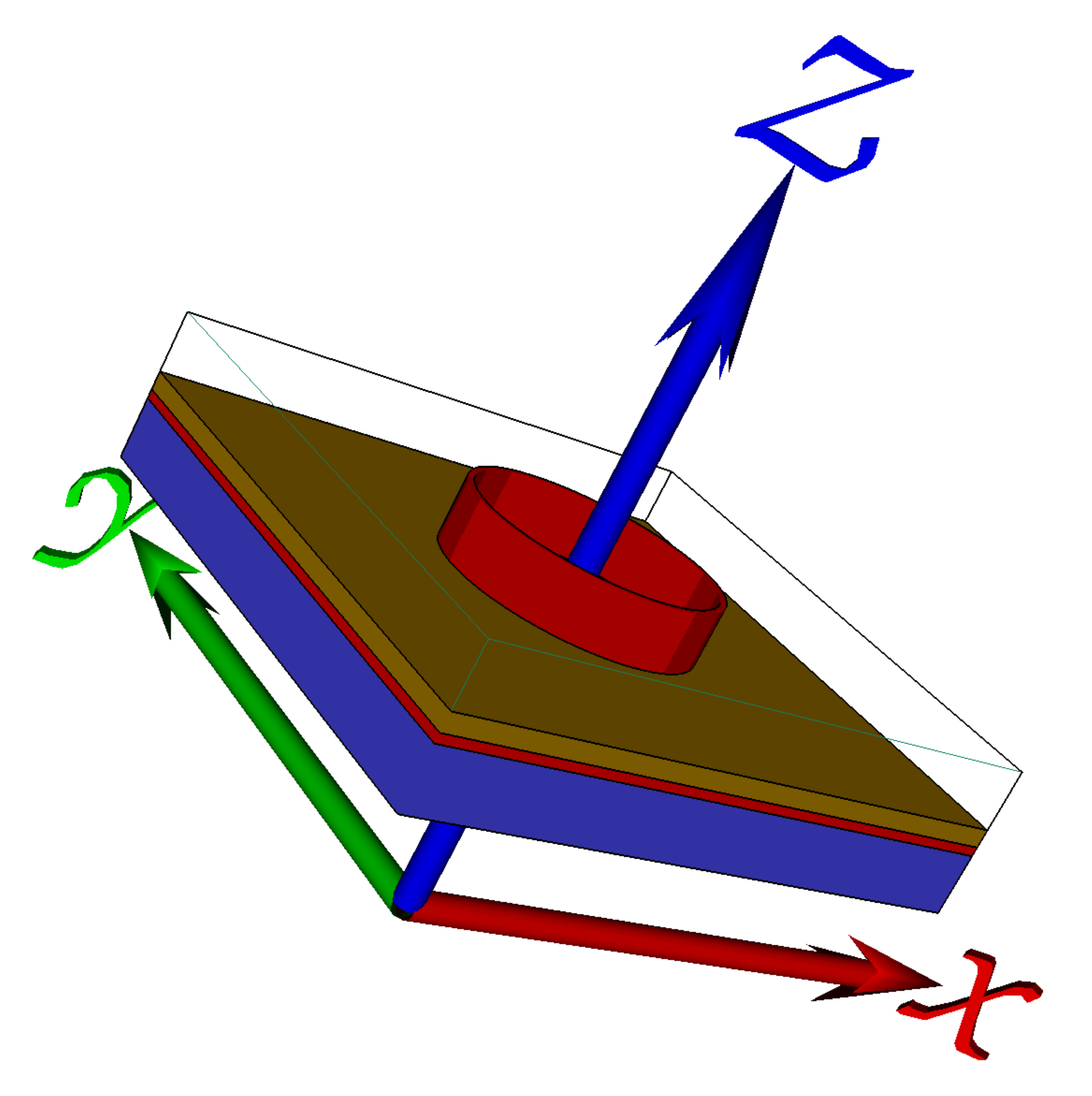}
    \caption{\label{fig:recovergeometry_geant4_with_barrier}Visualisation of the
      geometry as in figure \ref{fig:recovergeometry_geant4}, but with the
      back-scatter material opened up to show how a simple cylindrical barrier
      has been added inside, to contain the neutrons in a well-defined region.}
  \end{center}
\end{figure}

\begin{figure}[t!]
  \begin{center}
    \includegraphics[width=0.99\columnwidth]{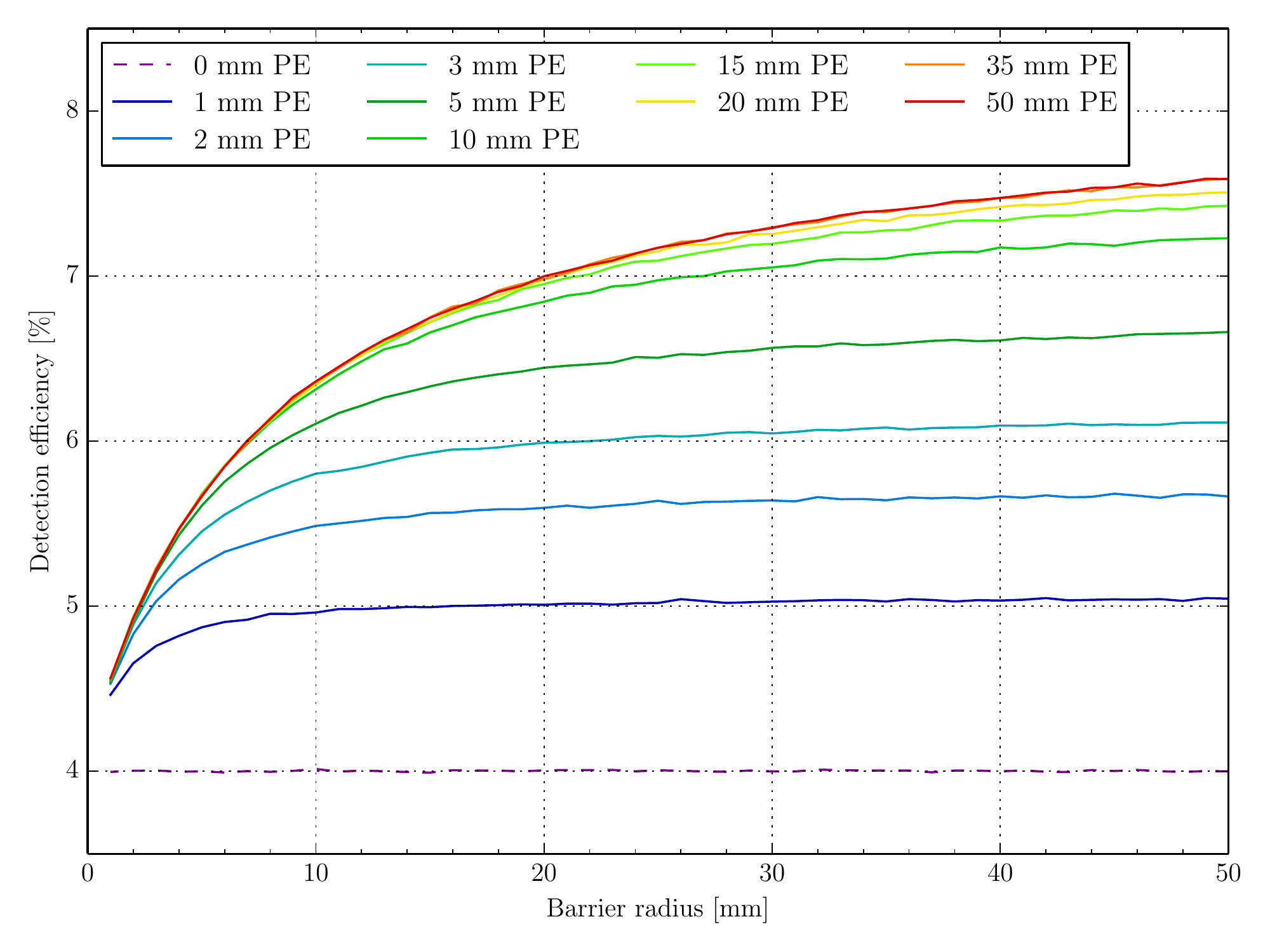}
    \caption{\label{fig:barrier_effect} Simulated detection efficiency for
      various amounts of polyethylene in the presence of a cylindrical barrier
      as a function of the barrier radius.}
  \end{center}
\end{figure}

\begin{figure}[t!]
  \begin{center}
    \includegraphics[width=0.99\columnwidth]{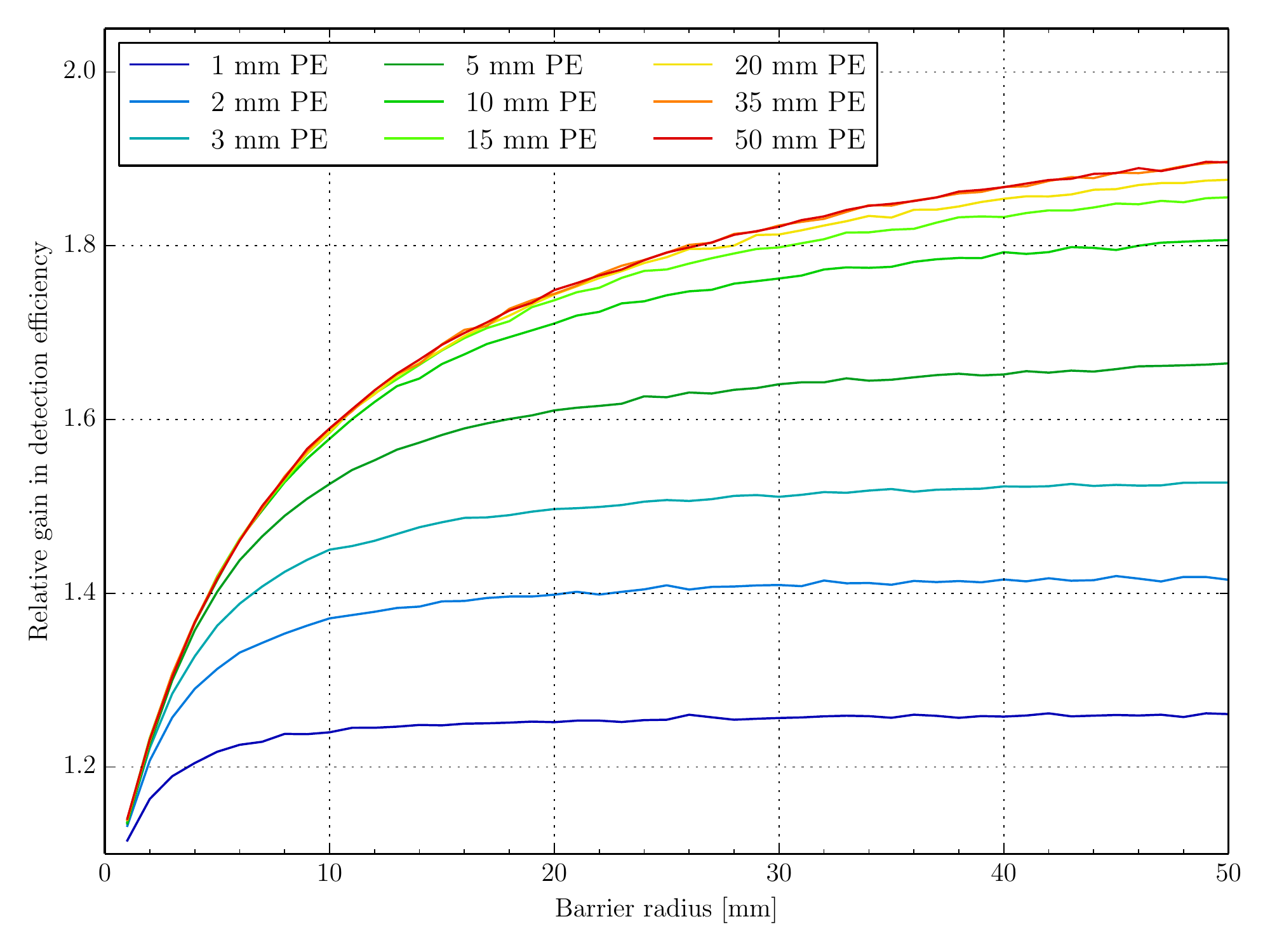}
    \caption{\label{fig:barrier_releffect} Simulated relative gain in detection
      efficiency for various amounts of polyethylene in the presence of a
      cylindrical barrier as a function of the barrier radius.}
  \end{center}
\end{figure}

Fortunately, tails in the spatial resolution are easily handled in the context
of real detectors, which are typically segmented according to their required
granularities, charge collection and readout schemes. Thus, by segmenting the
polyethylene as well, separating different detector cells by appropriate
absorbing material, one can be certain that a neutron entering the polyethylene
from within a given cell, will only be able to be back-scattered to the same
detection cell. The question then instead becomes one of how much the detection
efficiency gain due to the polyethylene will be reduced due to neutrons meeting
the absorber between polyethylene cells. This will again depend on the exact
detector design, but to quantify the effect and the dependency on the size of
detection cells, simulations were carried out with a cylindrical barrier of
100$\mu$m enriched boron-carbide placed in the polyethylene as shown in fig.\
\ref{fig:recovergeometry_geant4_with_barrier}, and detection efficiencies were
simulated for neutrons at normal incidence generated uniformly over the
corresponding circular detection cell. The result is shown in fig.\
\ref{fig:barrier_effect} and \ref{fig:barrier_releffect}: For very small
detection cells, the absorption in the barrier almost completely elliminates the
gain from adding the polyethylene, but for a realistic barrier radius of 20mm,
10mm of polyethylene will still provide a relative gain in detection efficiency
of approximately 70\%

\section{Conclusion}
The concept of enhancing effective detection efficiencies of neutron detectors
by placing a strongly scattering material at their backside was presented, and
investigated through analysis of \texttt{Geant4} simulations in the scenario of
polyethylene placed behind a single-layer thin-film detector. The method shows
great promise in the case of neutrons at low angle of incidence (close to
perpendicular to the Boron coating), but care must be taken to keep the
potential adverse consequences of the extra scatterings under control, possibly
via one of the investigated mitigation strategies.

\bibliography{refs_thki}

\vfill\newpage

\end{document}